\documentclass{article}
\usepackage{mathrsfs,array}
\usepackage{hyperref}
\usepackage{verbatim}
 \usepackage[T4]{fontenc}
\usepackage{amsfonts, amsthm, amssymb, amsmath}

\usepackage{graphicx}
\usepackage{subcaption}
\usepackage{qcircuit}
\usepackage{listings}
\usepackage{color}
\usepackage{cite}
\usepackage{subcaption}
\usepackage{authblk}
\definecolor{dkgreen}{rgb}{0,0.6,0}
\definecolor{gray}{rgb}{0.5,0.5,0.5}
\definecolor{mauve}{rgb}{0.58,0,0.82}

\usepackage[titletoc]{appendix}

\DeclareMathOperator\arctanh{arctanh}

\newcommand{\mb}{\mathbb}

\title{Approaching graph problems with continuous variable quantum computing}
\author[]{Micha{\l} St\k{e}ch{\l}y\footnote{michal.stechly@gmail.com}}
\author[]{Ntwali Bashige}
\author[]{Przemys{\l}aw Chojecki}
\affil[]{Bohr Technology Inc.}

\date{\today}

\begin{document}
\maketitle

\begin{abstract}

We introduce a method for solving the Max-Cut problem using a variational algorithm and a continuous-variables quantum computing approach. The quantum circuit consists of two parts: the first one embeds a graph into a circuit using the Takagi decomposition and the second is a variational circuit which solves the Max-Cut problem.
We analyze how the presence of different types of non-Gaussian gates influences the optimization process by performing numerical simulations. We also propose how to treat the circuit as a machine learning model.

\end{abstract}

\section{Introduction}

\noindent In recent years there has been a boom in quantum computing. Different models of hardware have been proposed, with the most dominant being based on superconducting circuits or trapped ions (cf. IBM, Rigetti, IonQ). All these architectures use discrete values (qubits) as the basis for computation. However, this is not the only possible approach --- one might also create a quantum computer based on qudits or even continuous variables \cite{gaussian_quantum_information}. The later is the approach that the Canadian company Xanadu is pursuing, building a photonic quantum computer utilizing the continuous variable paradigm. They have built the open-source libraries Strawberry Fields and PennyLane which provide tools for simulating photonic circuits and perfoming machine learning experiments.

\medskip 

\noindent Our goal in this paper is to take a hands-on approach towards optimization problems on continuous-variable quantum computers. This type of problem has already been researched both for discrete quantum computing and quantum annealing in \cite{qaoa}, \cite{crooks}, and \cite{maxcut_d_wave}.
\noindent In particular, we consider a solution to the Max-Cut problem on weighted graphs. We embed a graph into a quantum state, and then we optimize the parametrizable part of the circuit using a well-chosen cost function. For readers new to continuous-variable quantum computing, the appendices can serve as a quick introduction to the terminology and foundations of the field.

\subsection{Related work} 

In recent years, researchers started to use a new approach for creating quantum algorithms, namely variational circuits \cite{var_alg}. Algorithms like Variational Quantum Eigensolver (VQE) \cite{vqe} or Quantum Approximate Optimization Algorithm (QAOA) \cite{qaoa} have been succesfully employed to graph problems like the Max-Cut \cite{qaoa} and the traveling salesman problems \cite{hadfield}. However, these algorithms use the discrete variable quantum computing paradigm and there has not been much work done to solve this type of problems with continuous-variable quantum computing; with the notable recent exception of \cite{cv-qaoa}.
Among all the graph problems, the Max-Cut problem has thus far gotten much more attention than others as can be found in \cite{qaoa}, \cite{crooks}, and \cite{zhou}. Since many researchers use it as the first problem for testing and benchmarking their variational optimization algorithms, we decided to follow this trend.

\medskip

\noindent We have based our work on the methods described in \cite{gbs2} and \cite{cvqnn} and developed them further to solve the Max-Cut problem on a simulator of a photonic quantum computer.

\subsection{Acknowledgements}

We would like to thank Nathan Killoran and Josh Izaac for their help and guidance, Maria Schuld for help with the QMLT package and Nicol\'as Quesada for help with understanding the Takagi decomposition. We would also like to acknowledge Wayne Nixalo and Witold Kowalczyk for their contribution at the initial stage of this research.

\subsection{Organization}

This paper is organized as follows: in section \ref{sec:theoretical_framework} we present the theoretical framework we used in this research. We introduce two ways of representing a graph as a quantum state using either a Gaussian covariance matrix or the Takagi decomposition.

In section \ref{sec:experimental_results}, we present results of the simulation made on graphs of different sizes and offer preliminary results for extending the research to a machine learning model.

In section \ref{sec:conclusions}, we conclude the paper and give directions for further research.

Appendices \ref{sec:appendix_A}, \ref{sec:appendix_B}, \ref{sec:appendix_C} and \ref{sec:appendix_D} introduce the reader to the continuous-variables quantum computing paradigm.

\section{Theoretical framework}
\label{sec:theoretical_framework}

\subsection{Representation of a graph as a quantum state}

\noindent Let $(G,w)$ be a weighted graph where $G = (V,E)$ is a graph with vertices $V$ and edges $E$, and $w$ is a weight function which attaches to each edge between vertices $i$ and $j$ a real number $w(i,j)$. We set $w(i,j) = 0$ if there is no edge connecting $i$ and $j$.

\medskip

\noindent We write $A = (a_{ij})$ for the weighted adjacency matrix, that is we set $a_{ij} = w(i,j)$ for each $i,j \in V$. Inspired by \cite{gbs} and \cite{gbs2}, we match $A$ with a Gaussian covariance matrix (i.e. a matrix describing a state created by a Gaussian quantum circuit). We can either calculate it directly or perform the Takagi decomposition, which allows us to omit direct calculations. 
We describe both methods, but we use only the second one for the numerical simulations.

\subsubsection{Gaussian covariance matrix}

\noindent We assume that $G$ has $n$ vertices. 
Let $\mb{I}_n$ be the identity matrix. We define 

$$\mb{X} = \begin{bmatrix}0 & \mb{I}_n \\ \mb{I}_n & 0 \end{bmatrix}$$ 

\noindent The Gaussian covariance matrix associated with $A$ is defined to be
$$\sigma _{A} = (\mb{I}_{2n} - \mb{X}A)^{-1} - \mb{I}_{2n}/2$$
Let $c$ be an auxilary real number which will be our parameter to be determined for each graph separately. We choose $c$ such that $\sigma _{c\mb{I}_{2n}+A}$ is symplectic and positive definite, if that is possible. The reason we need to scale $A$ is because $\sigma_A$ does not always give a proper Gaussian covariance matrix\footnote{That is, it cannot be always represented by a gaussian quantum state} (cf. \cite{gbs2}, especially Appendix A). For arbitrary $A$, the above might not work. That is why we introduce $d$ to be another auxilary positive real, which we use as a parameter. Adopting the method in \cite{gbs2} we define
$$A' = \begin{bmatrix}A & 0 \\ 0 & A \end{bmatrix}$$
and set
$$\sigma ' _{c,d,A} = (\mb{I}_{4n} - \mb{X}d\cdot(c\mb{I}_{4n}+A'))^{-1} - \mb{I}_{4n}/2$$
where $\mb{X}$ here is defined with $\mb{I}_{2n}$. Pursuant to Appendix A in \cite{gbs2} for any $A$ there are always $c,d>0$ such that $\sigma ' _{c,d,A}$ is symplectic and positive definite.

\medskip

\noindent Now, using Strawberry Fields we are able to associate a quantum circuit with $\sigma _{d(c\mb{I}_{2n}+A)}$ if that is possible, or with $\sigma ' _{c,d,A'}$ if not. Note that we try to avoid using $\sigma ' _{c,d,A'}$ as it doubles the dimensions and the number of qumodes we need to use.

\noindent The result is a quantum circuit for which the output probability distribution depends on matrix $A$.

\noindent This method introduces additional parameters and requires choosing them in such a way that all the matrices meet the required conditions. However, the same result can be achieved by using the Takagi decomposition, as described in the section below. In this research we have used the later approach.

\subsubsection{Takagi decomposition}
\label{sec:takagi}
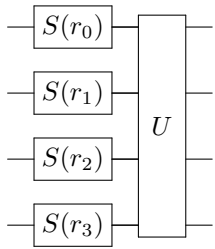
\begin{figure}
	\begin{center}
		\begin{minipage}{.7\textwidth}
			
			\Qcircuit @C=1em @R=.9em {
				& \gate{S(r_0)} & \multigate{3}{U}  \qw & \qw\\
				& \gate{S(r_1)} & \ghost{U}  \qw & \qw\\
				& \gate{S(r_2)} & \ghost{U}  \qw & \qw\\
				& \gate{S(r_3)} & \ghost{U}  \qw & \qw\\
			}
		\end{minipage}
	\end{center}
\caption{Circuit used for the preparation of the initial state. The parameters of the squeeze gates, as well as the exact form of the interferometer matrix come from the Takagi decomposition (see sec \ref{sec:takagi}).}
\label{fig:circuit_takagi}

\end{figure}

Let's take a set of $N$ squeezed states, with a squeezing parameter $r_i$, followed by an interferometer described by a matrix $U$ (see fig. \ref{fig:circuit_takagi}). If the matrices meet the condition:

\begin{equation}
\label{eq:takagi}
B = U D U^T
\end{equation}

\noindent where $D$ is a diagonal matrix with elements $d_i$ (which are the eigenvalues of $B$) on the diagonal and $r_i=\arctanh(d_i)$, then the probability distribution of such a state depends on matrix $B$ \cite{gbs}.

\medskip

\noindent Therefore, if we want to embed a weighted graph described by a distance matrix $B$ in a circuit, we do not need to calculate the covariance matrix; it's enough to perform the Takagi decomposition which is given by the equation \ref{eq:takagi} and to set the parameters of the gates accordingly. 

\medskip

\noindent There are two restrictions on matrix $B$: it has to be symmetrical and its eigenvalues must be from the interval $[-1, 1]$ so that it fits the $\arctanh$ function.
Matrix $A$ is always symmetrical, but it can have arbitrary eigenvalues. Therefore in order to embed it, we need to rescale it by multipling it by a constant, so that it meets the second condition.
\medskip

\noindent This method has several advantages over the previous one: it does not require calculating the covariance matrix explicitly, it does not introduce any parameters and it is much simpler. Those reasons make it an attractive method for state preparation as we do in this paper.

\subsection{Max-Cut problem}

Let $(G,w)$ be a weighted graph. A cut is a partition $(S,V \backslash S)$ of the vertex set $V$ into sets $S$ and $V \backslash S$. The weight $w(S, V \backslash S)$ of a cut is given by

\begin{equation}
\label{eq:1}
 w(S, V \backslash S) = \sum _{i \in S, j \in V \backslash S} w(i,j)
\end{equation}

\noindent The maximum cut is the cut of maximum weight and its weight is denoted by $mc(G,w)$, i.e.
$$ mc(G,w) = max _{S \subset V} w(S, V \backslash S)$$

\medskip

\noindent We can represent the set $S$ as a list: $s_N, s_{N-1}, ..., s_1$, where $N = |V|$ and $s_{i} \in {0, 1}$. With this representation we can express the weight of the cut $S$ as:

\begin{equation}
\label{eq:max_cut}
w(S) = \sum_{i,j \in E} w_{i,j} (1 - s_i * s_j),
\end{equation}

\noindent where $E$ is set of edges of the graph $(G,w)$ and $w_{i,j}$ is the weight of edge between nodes $i$ and $j$.

\medskip

\subsection{Circuit design}
\label{sec:circuit_design}

\begin{figure}
	\begin{center}
		\begin{minipage}{.7\textwidth}
			
			\Qcircuit @C=1em @R=.7em {
				& \gate{S(r_0)} & \multigate{3}{U} & \multigate{3}{U} & \gate{S} & \multigate{3}{U} & \gate{D} & \gate{NG} & \qw \\
				& \gate{S(r_1)} & \ghost{U} & \ghost{U} & \gate{S} & \ghost{U} & \gate{D} & \gate{NG} & \qw \\
				& \gate{S(r_2)} & \ghost{U} & \ghost{U} & \gate{S} & \ghost{U} & \gate{D} & \gate{NG} & \qw \\
				& \gate{S(r_3)} & \ghost{U} & \ghost{U} & \gate{S} & \ghost{U} & \gate{D} & \gate{NG} & \qw \\
			}
		\end{minipage}
	\end{center}
\caption{The circuit used for performing the optimization. The initial squeeze gates and the interferometer create the quantum state associated with the graph. Subsequent operations are parametrized (with the exception of the interferometers) and optimized (see sec. \ref{sec:circuit_design}).}
\label{fig:full_circuit}
\end{figure}
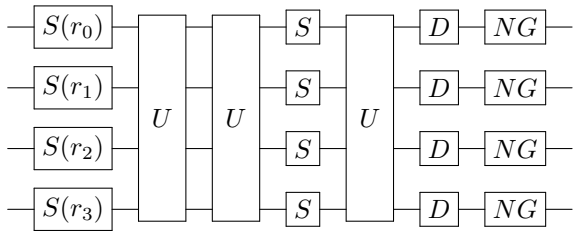

For simulating the circuits we used the Strawberry Fields library \cite{sf} and for training the parameters of those circuits we used the Quantum Machine Learning Toolbox (QMLT) \cite{qmlt}. Our circuit consists of two parts. The first one is associated with embedding the graph in the circuit, the second one is used for finding the solution for a given graph (see fig. \ref{fig:full_circuit}).
\medskip

\noindent We perform the embedding using the following procedure, according to sec. \ref{sec:takagi}:
 
\begin{enumerate}
	
	\item Create a distance matrix $A$ of the given graph.
	
	\item Rescale the matrix so all the eigenvalues are between -1 and 1 (excluding these values). After this procedure we get matrix $A'$.
	
	\item Perform the Takagi decomposition of the matrix $A'= U D U^T$.
	
	\item Take diagonal elements $d_i$ of the matrix $D$. The values $r_i = \arctanh(d_i)$ correspond to the initial squeezing of each mode.
	
	\item The matrix $U$ corresponds to the matrix describing an interferometer applied to the squeezed modes.
	
	\item The probability distribution of this state corresponds to the matrix $A$.
\end{enumerate}

\medskip

\noindent The second part of the circuit is based on the architecture proposed in \cite{cvqnn}. It consists of an interferometer, a layer of squeeze gates, a second interferometer, a layer of displacement gates and a layer of non-Gaussian gates --- either Kerr or cubic phase gates.

\medskip

\noindent Squeezing, displacement and non-Gaussian gates have been parametrized and the parameters of the interferometers have been fixed. We have done this in order to limit the number of parameters that need to be optimized and keep our analysis simple.

\medskip

\noindent All the parameters were initialized with random numbers. For both squeezing and displacement gates, the magnitude was drawn from the uniform distribution over [-0.5, 0.5] and the phase from the uniform distribution [0, 2$\pi$]. In case of non-Gaussian gates (which have only one parameter), it was also drawn from the uniform distribution over [-0.5, 0.5]. 

\medskip

\noindent Using a wider range as the support of the uniform distribution has been tested but it has not yielded any benefits in the results or training process and increased the risk of the simulation getting numerically unstable.

\subsection{Solution encoding}

We decided to use photon count as the output of the circuit, hence the output of each qumode could be in principle an integer from 0 up to infinity. However, in the simulation we are limited by the cutoff dimension so values for qumodes were capped at this value. In most cases it was equal to 17, but for some simulations we needed to lower this number to 9 due to memory constraints of the machines used. 

\medskip

\noindent In our case, binary encoding is sufficient since we divide nodes into two groups. This is why we decided to treat output of 0 as 0 and all non-zero outputs as ones. This means that the output [0, 5, 1, 3] will encode the solution [0, 1, 1, 1]. Other encodings are also possible, for example using homodyne measurements and encoding 0 and 1 in negative and positive values of position or momentum. 

\subsection{Training algorithm}
\label{sec:training}
For training the circuit we used the QMLT framework.
It has three modes: "optimization," "supervised" and "unsupervised". We have used "optimization," with stochastic gradient descent as an optimizer and L2-regularization. These are all standard choices --- something that deserves some more attention is the loss function we have used.

\medskip

\noindent For the classical, non-probabilistic algorithm, the most obvious choice of the loss function would be taking the output of the algorithm and calculating the cost from the equation \ref{eq:1}. However, in the case of the quantum circuit, the output is probabilistic, and so we need to take multiple results and calculate our loss function as an average over all the samples. A single result is a sample from a probability distribution hence with a growing number of samples we can reproduce the distribution more accurately.

\medskip

\noindent In our case, we are simulating the algorithm directly which enables us to use a probability distribution directly, without relying on the sampling. This could be approximated on the real devices by increasing the number of samples. The main problem with this approach is that in order to calculate the cost for the whole distribution, we also need to evaluate all the possible solutions classically. One might be inclined to ask what is the point of running the optimization procedure if we have to evaluate all the possible solutions classically anyway?

\medskip

\noindent There are two reasons why we do not think this is a significant issue in this case. Firstly, the problem we are dealing with is a toy problem and the research is preliminary. Hence the results we present can be treated as an upper bound on what could be done using this algorithm on a real machine instead of a simulator. Secondly, in the real scenario we will not be able to use the full distribution and we will need to rely on sampling. This will naturally limit the number of unique solutions we will need to evaluate.

\section{Experimental results}
\label{sec:experimental_results}

We have performed several tests to evaluate how the circuit works in different setups. In the spirit of other projects done by Xanadu, we used the codename "Yellow Submarine". The source code of the simulation can be found at: \url{https://github.com/BOHRTECHNOLOGY/yellow_submarine} and the code from the different experiments we performed at \url{https://github.com/BOHRTECHNOLOGY/public_research/tree/master/Experiments/Yellow_submarine}.

\subsection{Training parameters}
\label{sec:training_params}
We have been using the QMLT framework with initial learning rate equal to $0.25$ and regularization strength equal to $10^{-3}$.
The values of the regularization strength and the learning rate have been chosen experimentally. 
We have checked different values for these parameters and $10^{-3}$ was the highest value of regularization which had not forced the parameters to vanish over time.
Learning rates with values above $0.5$ resulted in instability during the learning process.

\subsection{Influence of the non-Gaussian gates}

We ran the simulation using both weighted and unweighted graphs with 4, 5 and 6 nodes. Of particular interest to us in the training is how the parameters of the different gates evolve, and more specifically, how non-Gaussian gates influence the simulation and its overall results. We investigate this for the displacement gate, the squeeze gate, the Kerr gate and the cubic phase gate. 
The plots we present here show the results for one 4-nodes graph, though they are representative for other graphs.

\subsubsection{Loss function}

We used a loss function as described in sec. \ref{sec:training}. In the case presented here, it could achieve a minimum value of -1 - this would mean that there is 100\% probability of getting a correct solution from the circuit. The second best solution had value of -0.75.

\medskip

\noindent As can be seen in the fig. \ref{fig:loss_function}, our circuits converged to value around -0.9, which indicates that the correct solution was the most frequent one. Also, the results are similar regardless of the non-Gaussian gates used. 

\medskip

\noindent Solutions of the Max-Cut problem have symmetry: [1, 0, 0, 1] has the same cost as [0, 1, 1, 0]. It is worth noting that training always converged to returning only the single best solution, not a superposition of all the best solutions.

\medskip

\subsubsection{Displacement gate parameters}

The displacement gate has two parameters: magnitude and phase. Values of magnitude always converged towards one of two or three values (see fig. \ref{fig:d_gate} A). This suggests that during the learning process, the displacement gate parameter magnitude contributes significantly to the end result --- this has also been confirmed by a simulation with the displacement gates removed from the circuit.

\medskip

\noindent We also note that the phase parameter usually does not change much from the initial value (see fig. \ref{fig:d_gate} B), and the rate of change is much smaller than in the case of magnitude. This suggests that phase paramater plays less significant role than the magnitude part. 

\medskip
\subsubsection{Squeeze gate parameteres}

The squeeze gate parameter magnitude does vary during the simulation but does not converge towards any specific value (see fig. \ref{fig:s_gate}). The parameter phase vary only a little bit throughout the simulations. Therefore, we conclude that as with the displacement gate, the phase parameter is less important than the magnitude.

\medskip

\subsubsection{Kerr gate}

The Kerr gate parameter remains entirely constant throughout the entire training for all simulations. The small change in the value of this parameter that can be seen in fig. \ref{fig:kerr_gate} comes from regularization.

This, in conjunction with the fact that the results with and without Kerr gate are very similar, seems to indicate that the Kerr gate does not participate at all in the computation of the final answer.

\medskip
\subsubsection{Cubic phase gate}

The cubic phase gate parameter does change during the training, but its behavior is much less consistent. Most often it just changes during the training, but sometimes converges towards specific values (see fig. \ref{fig:cubic_phase_gate}). It also does not speed up the convergence or help to lower the final value of the cost function.
Additionally, The presence of the cubic phase gate sometimes induced spikes in the loss function which shows the instability it introduces in the training process. This needs to be compensated for with different hyper-parameters and might be the subject of a more comprehensive study in the future.

\subsection{Influence of the embedding}

In the setup that we have proposed, we can omit the embedding part of the circuit and use only the variational part. This means that the variational part will act on a vacuum state instead of state corresponding to a graph. Since information about the graph structure is encoded in the cost function, the optimization process will still drive the solution toward some local minimum. We have checked whether the presence of the graph embedding improves the results.

\medskip

\noindent Depending on the graph we tried to solve, the presence of the embedding had negligible to slightly negative influence on the results and training process. 
It was the strongest for the graph with 4 nodes, where the final value of the cost function was on average 10\% higher and the convergence was up to two times slower in some cases. However, this effect was much weaker for graphs with 5 and 6 nodes, sometimes even unobservable.

\medskip

\noindent We have checked if circuits containing some non-Gaussian gates are influenced more than the others, but no such correlation has been found.

\subsection{A quantum circuit as a machine learning model}

We also wanted to check if our circuit can be treated as a machine learning model, i.e. if it can be trained using one set of graphs and then generalize to solve graphs that had not been presented to it.

\medskip

\noindent We used the following procedure to achieve this:

\medskip

\noindent Given a training set $X$ of $n$ matrices $x_i$, for each optimization step we have embedded every matrix $x_i$ once, and calculated the loss function. Then we have taken the sum of the loss functions for the whole set $X$ and used it to update the values of the parameters (in order to achieve this, we needed to slightly modify the source code of the QMLT library. The source code is available in the repository).

\medskip

\noindent In our case the set $X$ consisted of four 4x4 matrices. Each matrix represented a graph with star topology --- there was a central node and all the other nodes were connected only to it. In each set of the matrices, a different node was the central one. This type of a training set has important properties:

\begin{itemize}
\item A star topology guarantees that we will always have only one optimal solution - namely that the central node is in one group and all the other nodes are in the other one. 
\item Since in each case the central node is different, all the graphs in the set $X$ have different optimal solutions.
\item All the possible solutions occur in equal proportion. %TODO: with the same probability?
\item It is therefore easy to tell if a given circuit has learned to solve the problem for a given graph, or if it simply converged to one of the local minima. 
\end{itemize}

\medskip

\begin{figure}[ht]
	\centering
	\includegraphics[width=50mm]{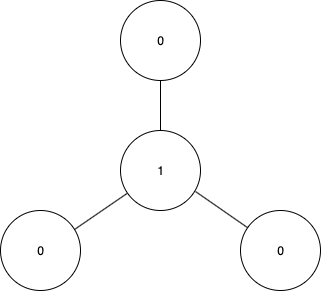}
	\caption{The star graph used in the "machine learning model" approach. The numbers indicate assignment to one of the groups. The assignment shown in the figure is the optimal solution to the Max-Cut problem for this graph. For each graph in the training set X, the central nodes had different indices.}
\label{fig:star_topology}
\end{figure}

\medskip

\noindent Apart from the original architecture, we have also tested another setup, where the parametrized part was duplicated - hence we had two layers of gates. In both cases circuits failed to solve the problem correctly. In the end they learned to return the same output regardless of what graph had been embedded in the circuit. The outputs varied between different runs, but they always converged to a configuration where two bits were on and two were off. For different non-Gaussian gates used, there was no difference in the quality of the output. However, the training process without the non-Gaussian gates was much smoother while adding non-Gaussian gates introduced oscillations, with amplitude varying between different runs (see fig. \ref{fig:ml_results}).

\section{Conclusions}
\label{sec:conclusions}

\noindent In this study, we have created a framework for solving the Max-Cut problem using photonic quantum circuits. We have checked its performance for graphs of up to 6 nodes and we have checked how using different gates affects the training process of parametric circuits. Importantly, we compared the performance of this optimization method in a scenario where only  Gaussian gates are used versus one where non-Gaussian gates are added.

\medskip

\noindent Based on the results of the numerical simulations we can say that:

\begin{itemize}
\item the setup that we proposed allows to solve Max-Cut problem.
\item the presence of non-Gaussian gates does not yield any improvement and might even result in instabilities in the optimization process.
\item starting from a state described in \ref{sec:takagi} might have detrimental effect on the results.
\end{itemize}

\noindent Since the work we have done was mostly experimental, we think that these conclusions are not definite, but might be useful for other researchers implementing a variational algorithm in a continuous-variable quantum computing model. Experimenting with the machine learning approach seems especially promising, since we have only touched this topic.

\noindent Also the fact that non-Gaussian are not needed to solve the Max-Cut problem seems interesting. This suggests that Gaussian Boson Sampler, which is a device that is simpler to build than a full continuous-variable quantum computer, might be useful for solving graph problems. On the other hand, it is unclear whether this approach gives any advantage in scaling or performance over classical methods --- it would require further investigation.

\medskip

\noindent We invite other researchers to use our code. Links to the code have been provided in section \ref{sec:experimental_results}. The natural next step is to look at other combinatorial optimization problems like the Traveling Salesman Problem. We are aware that during the work on this project new tools have been released, like the PennyLane library or a new version of Strawberry Fields, but we nevertheless think that having access to the source code might be helpful.

\newpage

\appendix
\section{Introduction to continuous-variable quantum computing}
\label{sec:appendix_A}

\noindent Most of the attempts to build a universal quantum computer have been
directed towards creating a discrete quantum computer that uses two
quantum bits in a similar vein to a classical computer using two classical
bits. In this approach, the quantum equivalents of bits are implemented
using discrete quantum phenomena such as electron spin. This scheme
is called discrete-variable quantum computing.

\medskip

\noindent In continuous-variable quantum computing we are not restricted
to discrete values but can have continuous values as well. For the
purpose of this paper, we define the state of a continuous-variable
quantum computer and some operations on those states.

\subsection{Qumodes}

In classical computing, the basic computing unit is the bit. In discrete
variables quantum computing, it is the qubit. In continuous-variable
quantum computing the \emph{qumode} is the basic computing unit.

\medskip

\noindent As a physical system, we can use a quantum harmonic oscillator as
a model that allows us to investigate the state of a continuous-variable quantum system and its associated qumodes. In general, for
each state of the system, there is an associated qumode. Therefore,
we make no distinction between state and qumode when there is no risk
of confusion.

\medskip

\noindent Mathematically if this system has a Hamiltonian $\hat{H}$ then it
is described by the following:
\begin{equation}
\hat{H}=\sum_{k=1}^{N}\hat{H}_{k}
\end{equation}
\noindent The Hamiltonian $\hat{H_{k}}$ denotes the Hamiltonian of the $k^{th}$
harmonic oscillator among $N$ qumodes.

\medskip

\noindent From elementary quantum mechanics, we know that the Hamiltonian of
the quantum harmonic oscillator is given by (assuming unit mass):
\begin{equation}
\hat{H}_{k}=\frac{1}{2}\omega_{k}^{2}\hat{x}_{k}^{2}+\frac{\hat{p}_{k}}{2}
\end{equation}
\noindent In equation $(2)$, $\omega_{k}$ is the frequency of the $k^{th}$
mode. The position $\hat{x}_{k}$ and momentum $\hat{p}_{k}$ operators
of the $k^{th}$ mode obey the commutation relationship $\left[\hat{x}_{k},\hat{p}_{k}\right]=i\hslash$.

\medskip

\noindent We introduce two new sets of operators defined in terms of $\hat{x}_{k}$
and $\hat{p}_{k}$:

Quadratures of a single qumode $k$ are defined as follows
\begin{equation}
\hat{X}_{k}=\sqrt{\frac{\omega_{k}}{\hslash}}\hat{x}_{k}
\end{equation}
\begin{equation}
\hat{P}_{k}=\sqrt{\frac{1}{\hslash\omega_{k}}\hat{p}_{k}}
\end{equation}
The Hamiltonian can be defined in terms of these quadratures:
\begin{equation}
\hat{H}_{k}=\frac{\hslash\omega}{2}\left[\hat{X}_{k}^2+\hat{P}_{k}^2\right]
\end{equation}

\noindent The quadrature operators obey the commutation relation $\left[\hat{X}_{m},\hat{P}_{n}\right]=i$.
Since they are Hermitian, they correspond to physical observables
that can be measured.

\subsection{Vacuum state}

There exists a special state that corresponds the lowest-energy state of
the quantum harmonic oscillator. This state is the vacuum state (also
called the ground state). This state is important because coming from
classical mechanics, we assume that the vacuum has no energy but in
quantum mechanics the vacuum actually has energy and we must take that
into account in continuous-variable quantum computing. This is best
demonstrated in the analysis of a beamsplitter: if we send a photon
in one port of the beamsplitter but not the other port, our analysis
must treat the unused port as vacuum and if we require the use of
the Hamiltonian of either port, the Hamiltonian of the unused port
shall not be zero but will be constrained by the Heinsenberg uncertainty
principle.

\section{Gaussian states and operators}
\label{sec:appendix_B}

\noindent A certain class of states, the so-called Gaussian states, are extremely
important because they are easy to efficiently produce in the laboratory.
This is important because while we know that Gaussian states are not
enough for universal computation, we still want to investigate what
we can accomplish with them alone that we cannot accomplish classically.

\medskip

\noindent In this section we introduce and describe Gaussian states and Gaussian
operators. We begin by introducing Gaussian functions then proceed
to the Wigner distribution and finally we elaborate on Gaussian states
and operators.

\subsection{Gaussian functions}

Found under the name of normal distributions in mathematics literature,
Gaussian functions occur with striking regularity throughout science.
We have already encountered a phenomena whose state is described by
a Gaussian function: the vacuum state.

\medskip

\noindent A multi-variate Gaussian function has the form
\begin{equation}
G(x)=C\times exp\left\{ -\frac{1}{2}x^{T}Ax+b^{T}x\right\} 
\end{equation}
where $x=\left(x_{1},x_{2},...,x_{N}\right)^{T}$, $b=(b_{1},b_{2},...,b_{N})^{T}$
and $A$ is a $N\times X$ positive-definite matrix.

\medskip

\noindent We do not delve into the details of Gaussian functions but we invite
the reader to keep the form of the function in their minds.

\subsection{Wigner functions}

In 1932 in a paper titled ``On the quantum correction for thermodynamics
equilibrium'', Eugene Wigner formulated a function which is a \emph{quasi-probability
	distribution} for multiple particles. For a single particle in the
$x$ basis, the Wigner function has the following form:

\begin{equation}
W(x,y)=\frac{2}{\pi}\int_{-\infty}^{+\infty}dy\times exp\left\{ +4iyp\right\} \times\langle x-y\vert\hat{\rho}\vert x+y\rangle
\end{equation}

\noindent We shall not solve this function for particular quantum systems but
the reader is invited to understand what it means: the Wigner function
is a quasi-probability distribution that describes the effects on
the quadratures. If the solution to the Wigner function for a particular
quantum state (system) in a given basis has the form of a Gaussian
function, we say that such a state (system) is a Gaussian state (system).
This function belongs to a wider class of functions called \emph{quasi-probability
	distributions} because while it is normalized, gives the correct marginal
distributions and allows the calculations of averages and variances
of quadratures, it can return negative values which are not valid
probability values.

\subsection{Gaussian states}

Gaussian states occur in three main ways: as pure coherent states,
pure squeezed states or squeezed coherent states.
\begin{enumerate}
	\item Coherent states are states of minimum uncertainty equally dispersed
	among the quadratures. They are created by applying the displacement
	operator to a qumode in the vacuum state. This means that the vacuum
	state is also a coherent state because it can be obtained by applying
	the displacement operator to the vacuum with zero displacement.
	\item Squeezed states refer to states where the quantum fluctuations in
	one quadrature are reduced at the expense of an increased uncertainty
	in the conjugate quadrature. A squeezed state is created by applying
	the squeezing operator to a qumode. The squeezing operator shall be
	discussed in the next section.
	\item Squeezed coherent states refer to states that have the displacement
	operator applied to the vacuum followed by the application of the
	qumode.
\end{enumerate}

\subsection{Gaussian operators}

An operator is Gaussian if it transforms a Gaussian state into another
Gaussian state. We mentioned two Gaussian operators (displacement and squeezing) in the preceding
section, now we elaborate on them and we introduce two more: the single-mode rotation operator and the two-modes beamsplitter.

\subsubsection{The displacement operator}

The displacement operator $\hat{D}(\alpha)$ increases the measured
position by $Re(\alpha)$ and the measured momentum by $Im(\alpha)$
for $\alpha\in\mathbb{C}$. Its action on the position quadrature
is given by

\begin{equation}
\hat{D}^{\dagger}(\alpha)\hat{x}\hat{D}(\alpha)=\hat{x}+\sqrt{2\hslash}Re(\alpha)I
\end{equation}

Its action on the momentum quadrature is given by

\begin{equation}
\hat{D}^{\dagger}(\alpha)\hat{p}\hat{D}(\alpha)=\hat{p}+\sqrt{2\hslash}Im(\alpha)I
\end{equation}

In matrix form, taking $\hslash=\frac{1}{2}$ and replacing the
identity matrix with $1$, the displacement operator acts as follows

\begin{equation}
\begin{bmatrix}x'\\
p'
\end{bmatrix}=\begin{bmatrix}x+Re(\alpha)\\
p+Im(\alpha)
\end{bmatrix}
\end{equation}

Where $\begin{bmatrix}x & p\end{bmatrix}^{T}$ is the real-value vector
of the measured position and momentum.

\subsubsection{The squeezing operator}
\label{app:squeeze}

The squeezing operator $\hat{S}(r)$with squeezing parameter $r$
reduces the uncertainty in either position or momentum (well below
the standard quantum limit) while increasing the uncertainty in the
conjugate variable. Its action on the position quadrature is given
by
\begin{equation}
\hat{S}^{\dagger}(r)\hat{x}\hat{S(r)}=e^{-r}\hat{x}
\end{equation}

Its action on the momentum quadrature is given by

\begin{equation}
\hat{S}^{\dagger}(r)\hat{p}\hat{S(r)}=e^{+r}\hat{p}
\end{equation}

In matrix form, the squeezing operator acts as follows
\begin{equation}
\begin{bmatrix}x'\\
p'
\end{bmatrix}=\begin{bmatrix}e^{-r} & 0\\
0 & e^{+r}
\end{bmatrix}\begin{bmatrix}x\\
p
\end{bmatrix}
\end{equation}

Where $\begin{bmatrix}x & p\end{bmatrix}^{T}$ is the real-value vector
of the measured position and momentum.

\subsubsection{The rotation operator}

The rotation operator $R(\phi)$ rotates quadratures in phase space
by $\phi\in\left[0,2\pi\right]$. Its action on the position quadrature
is
\begin{equation}
\hat{R}^{\dagger}(\phi)\hat{x}\hat{R}(\phi)=\hat{x}\cos\phi-\hat{p}\sin\phi
\end{equation}

Its action on the momentum quadrature is
\begin{equation}
\hat{R}^{\dagger}(\phi)\hat{p}\hat{R}(\phi)=\hat{x}\sin\phi+\hat{p}\cos\phi
\end{equation}

In matrix form, the rotation operator acts on position and momentum
as follows

\begin{equation}
\begin{bmatrix}x'\\
p'
\end{bmatrix}=\begin{bmatrix}\cos\phi & \sin\phi\\
-\sin\phi & \cos\phi
\end{bmatrix}\begin{bmatrix}x\\
p
\end{bmatrix}
\end{equation}
Where $\begin{bmatrix}x & p\end{bmatrix}^{T}$ is the real-value
vector of the measured position and momentum.

\subsubsection{The beamsplitter operator}

The beamsplitter operator $\hat{BS}(\theta,\phi)$ is a two-mode operator
that requires two qumodes to operates upon. It may be understood as
a rotation between two qumodes. Its action on the position quadratures
is as follows

\begin{equation}
\begin{cases}
\hat{BS}^{\dagger}(\theta,\phi)\hat{x}_{1}\hat{BS}(\theta,\phi) & =\hat{x}_{1}\cos\theta-\sin\theta\left[\hat{x}_{2}\cos\phi+\hat{p}_{2}\sin\phi\right]\\
\hat{BS}^{\dagger}(\theta,\phi)\hat{x}_{2}\hat{BS}(\theta,\phi) & =\hat{x}_{2}\cos\theta+\sin\theta\left[\hat{x}_{1}\cos\phi-\hat{p}_{1}\sin\phi\right]
\end{cases}
\end{equation}

Its action on the momentum quadratures is given by
\begin{equation}
\begin{cases}
\hat{BS}^{\dagger}(\theta,\phi)\hat{p}_{1}\hat{BS}(\theta,\phi) & =\hat{p}_{1}\cos\theta-\sin\theta\left[\hat{p}_{2}\cos\phi+\hat{x}_{2}\sin\phi\right]\\
\hat{BS}^{\dagger}(\theta,\phi)\hat{p}_{2}\hat{BS}(\theta,\phi) & =\hat{p}_{2}\cos\theta+\sin\theta\left[\hat{p}_{1}\cos\phi-\hat{x}_{1}\sin\phi\right]
\end{cases}
\end{equation}

In matrix form, the beamsplitter acts as follows

\begin{equation}
\begin{bmatrix}x'_{1}\\
x'_{2}\\
p'_{1}\\
p'_{2}
\end{bmatrix}=\begin{bmatrix}\cos\theta & -\sin\theta\cos\phi & 0 & \sin\theta\sin\phi\\
\sin\theta\cos\phi & \cos\phi & -\sin\phi\cos\theta & 0\\
0 & \sin\phi\cos\theta & -\cos\theta & -\sin\theta\cos\phi\\
-\sin\phi\cos\theta & 0 & \sin\theta\cos\phi & \cos\theta
\end{bmatrix}\begin{bmatrix}x_{1}\\
x_{2}\\
p_{1}\\
p_{2}
\end{bmatrix}
\end{equation}

Where $\begin{bmatrix}x_{1} & x_{2} & p_{1} & p_{2}\end{bmatrix}^{T}$
is the real-value vector of the measured positions and momenta.

\section{Non-Gaussian states and operators}
\label{sec:appendix_C}

\noindent We will not have much to say about non-Gaussian states nor about non-Gaussian
operators but mention two non-Gaussian operators and technical challenges
around their implementation.

\subsection{Non-Gaussian states}

Non-Gaussian states are simply ones where the Wigner function does not
results in a Gaussian function. They can be created when non-Gaussian
operators act upon Gaussian states.

\subsection{Non-Gaussian operators}

Non-Gaussian operators are ones that do not return a Gaussian state
when given a Gaussian state. Commonly used non-Gaussian operators
are the Kerr operator and the Cubic phase operator.

\subsubsection{Cubic phase operator}

The cubic phase operator $V(\gamma)$ can be understood as the result
of the following process: prepare two qumodes in the vacuum state.
Apply a two-mode squeeze operator on both qumodes. Then apply a large
momentum displacement on the first qumode. Lastly, perform a photon counting
measurement on the first qumode. This results in the second qumode
behaving as if the cubic phase operator was applied to it and its
state may be called a cubic phase state which is a non-Gaussian state.

\medskip

\noindent Its action on the position quadrature is given by
\begin{equation}
\hat{V}(\gamma)\hat{x}\hat{V}(\gamma)=\hat{x}
\end{equation}

\noindent Its action on the momentum quadrature is given by
\begin{equation}
\hat{V}(\gamma)\hat{p}\hat{V}(\gamma)=\hat{p}+\gamma\hat{x}
\end{equation}

\subsubsection{Kerr operator}

Let $\hat{n}_{k}=\frac{1}{2}\left[\hat{X}_{k}^2+\hat{P}_{k}^2\right]$ be defined in terms of quadratures. $\hat{n}$ is called the number operator. We then proceed to define the Kerr operator with reference to the number operator with parameter $\kappa$.

\begin{equation}
K = exp\left\{ i \kappa \hat{n}^2\right\}
\end{equation}

\section{Measurements}
\label{sec:appendix_D}

\noindent In continuous-variable quantum computing, we have an array of measurements
we can choose from: photon counting, homodyne detection and heterodyne
detection.

\subsection{Photon counting}

Photon counting measurement projects the qumode onto the number eigenstates $\vert n\rangle$ resulting in a natural number $n$.

\medskip

\noindent For $n = 0$, photon counting of a single	qumode in a multimode Gaussian state preverves the Gaussian character of the remaining qumodes. The same is not true for $n \textgreater 0$.

\subsection{Homodyne detection}

In homodyne detection, photons are captured by a photodetector which
converts photons into electrons resulting in an electric current.
This electric current is generally proportional to the number of photons.
We wish to detect a quadrature of the qumode under consideration.
To accomplish this, the qumode is combined with a 50/50 beam splitter.
The two output qumodes of the beam splitter are the ones that are converted
to photocurrents and their difference measured. The quadrature of
choice is measured by introducing a phase in the beam splitter which
is varied to select the quadrature.

\medskip

\noindent Homodyne detection is a projective Gaussian measurement that projects
Gaussian states onto other Gaussian states.

\subsection{Heterodyne detection}

Heterodyne measurement is a measurement of both the $\hat{x}$ and $\hat{p}$ quadratures simultaneously. This results in uncertainty in the measurement of quadratures.

\medskip

\noindent As with homodyne measurement, heterodyne measure projects Gaussian states onto other Gaussian states.

\newpage

\begin{figure}
\centering
\begin{tabular}{@{}c@{}}
    \includegraphics[width=0.9\linewidth,height=130pt]{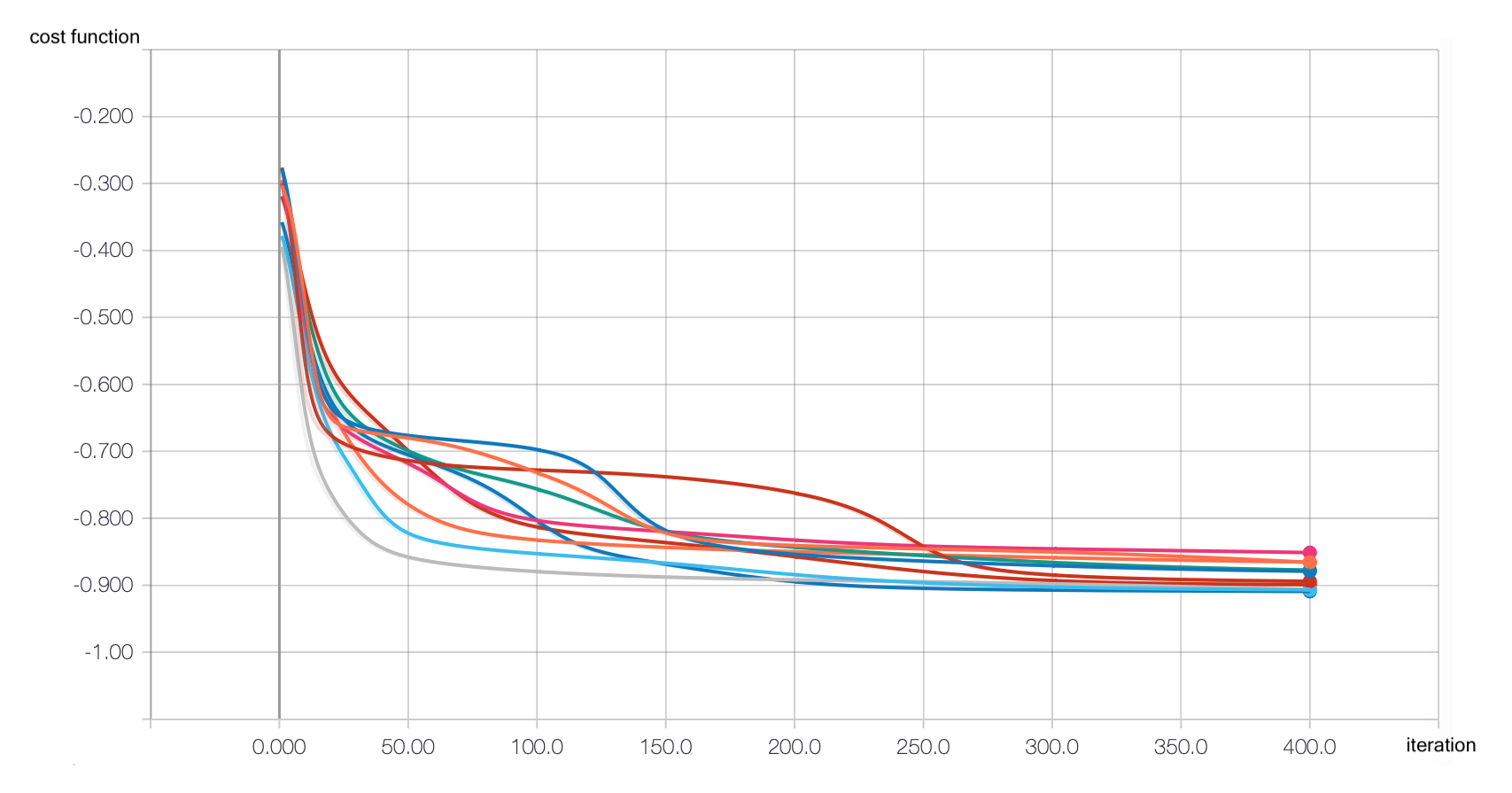} \\[\abovecaptionskip]
    \small A
\end{tabular}

\vspace{\floatsep}

\begin{tabular}{@{}c@{}}
    \includegraphics[width=0.9\linewidth,height=130pt]{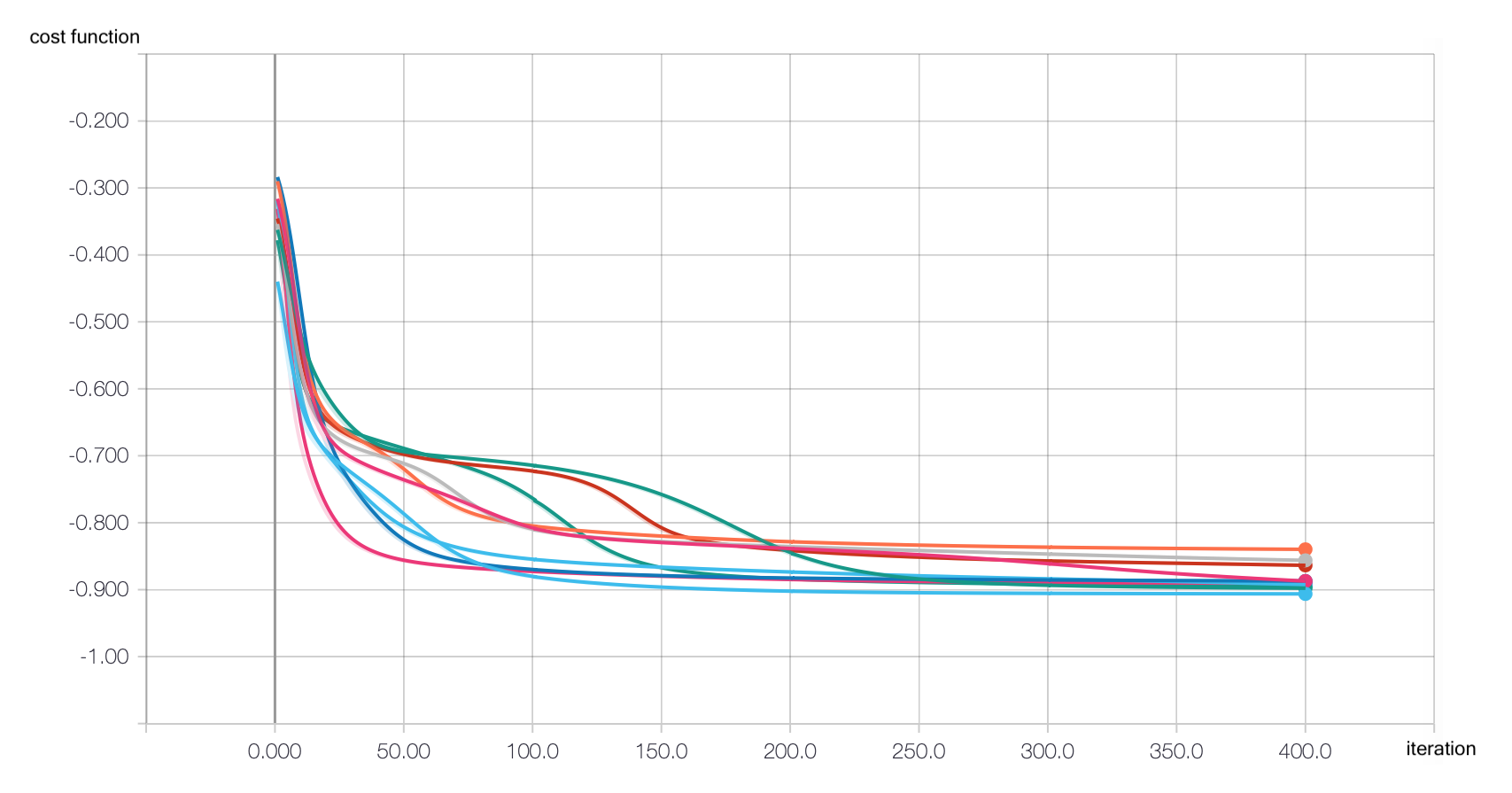} \\[\abovecaptionskip]
    \small B
\end{tabular}

\begin{tabular}{@{}c@{}}
    \includegraphics[width=0.9\linewidth,height=130pt]{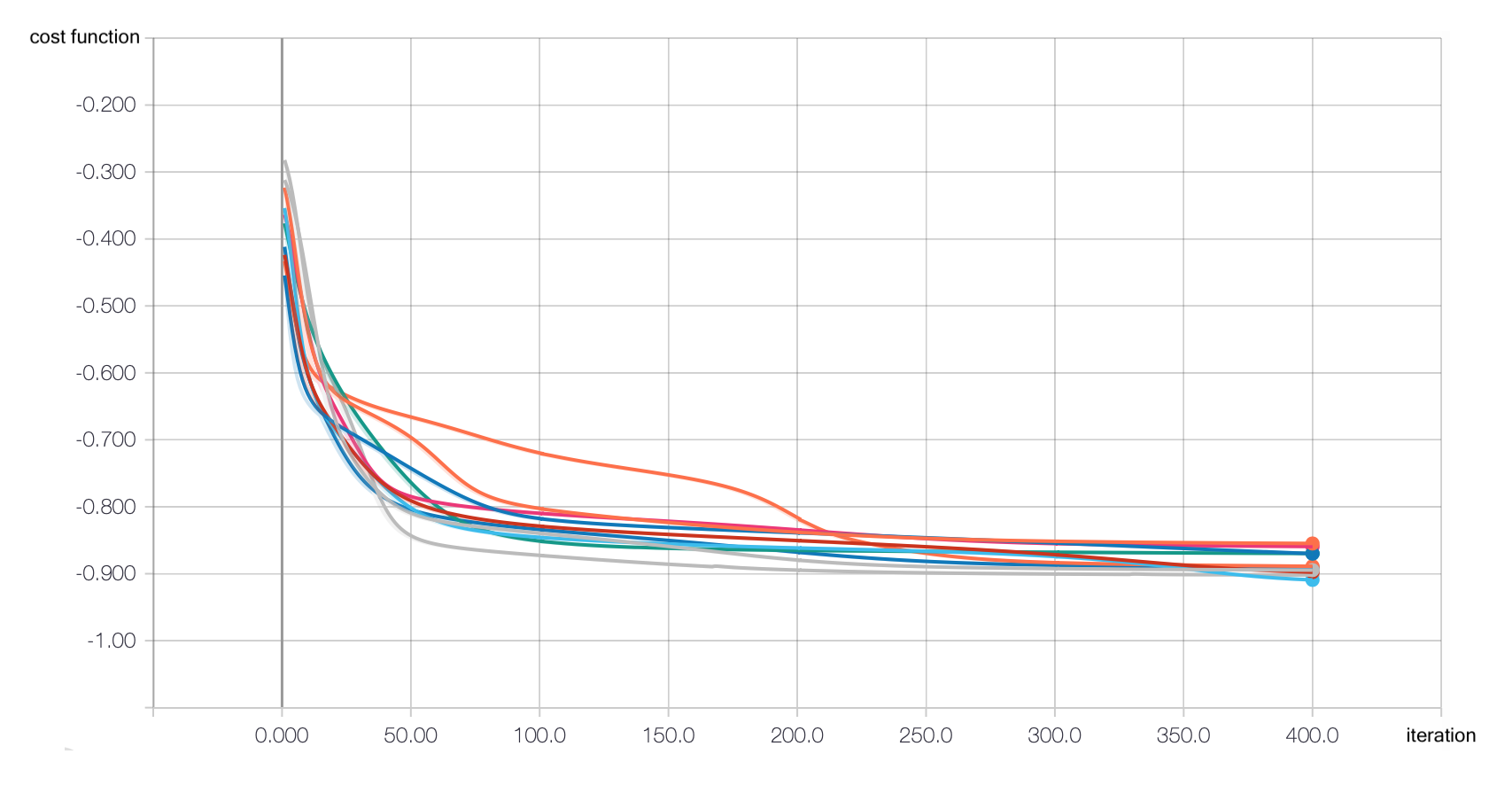} \\[\abovecaptionskip]
    \small C
\end{tabular}

\caption{The plots of the loss function for several simulations without any non-Gaussian gates (A), Kerr gates (B) and cubic phase gates (C). The minimum possible value of the loss function is -1.}
\label{fig:loss_function}
\end{figure}

\begin{figure}
    \centering
    \begin{tabular}{@{}c@{}}
        \includegraphics[width=1\linewidth,height=150pt]{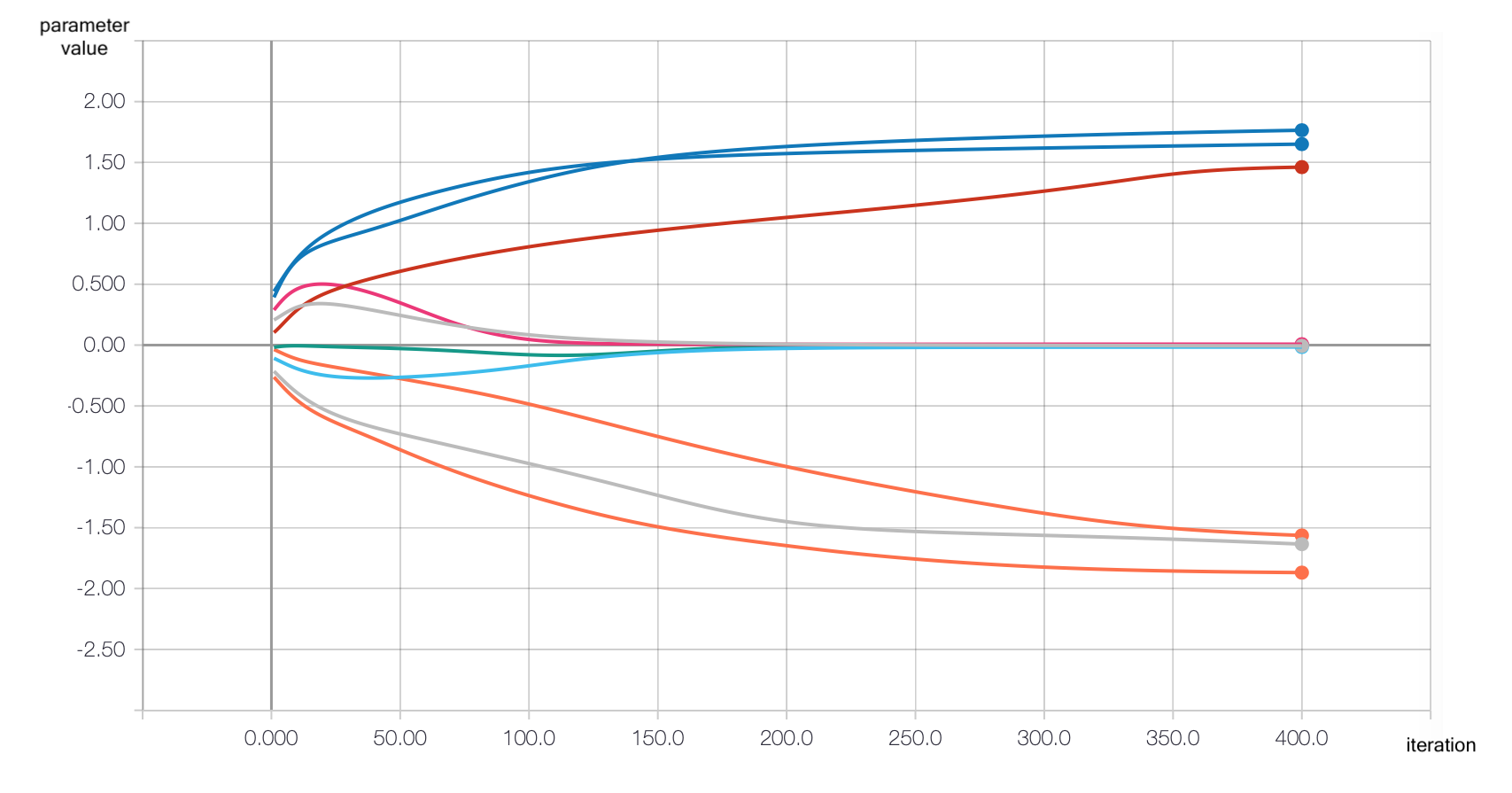} \\[\abovecaptionskip]
        \small A
    \end{tabular}
    
    \vspace{\floatsep}
    
    \begin{tabular}{@{}c@{}}
        \includegraphics[width=1\linewidth,height=150pt]{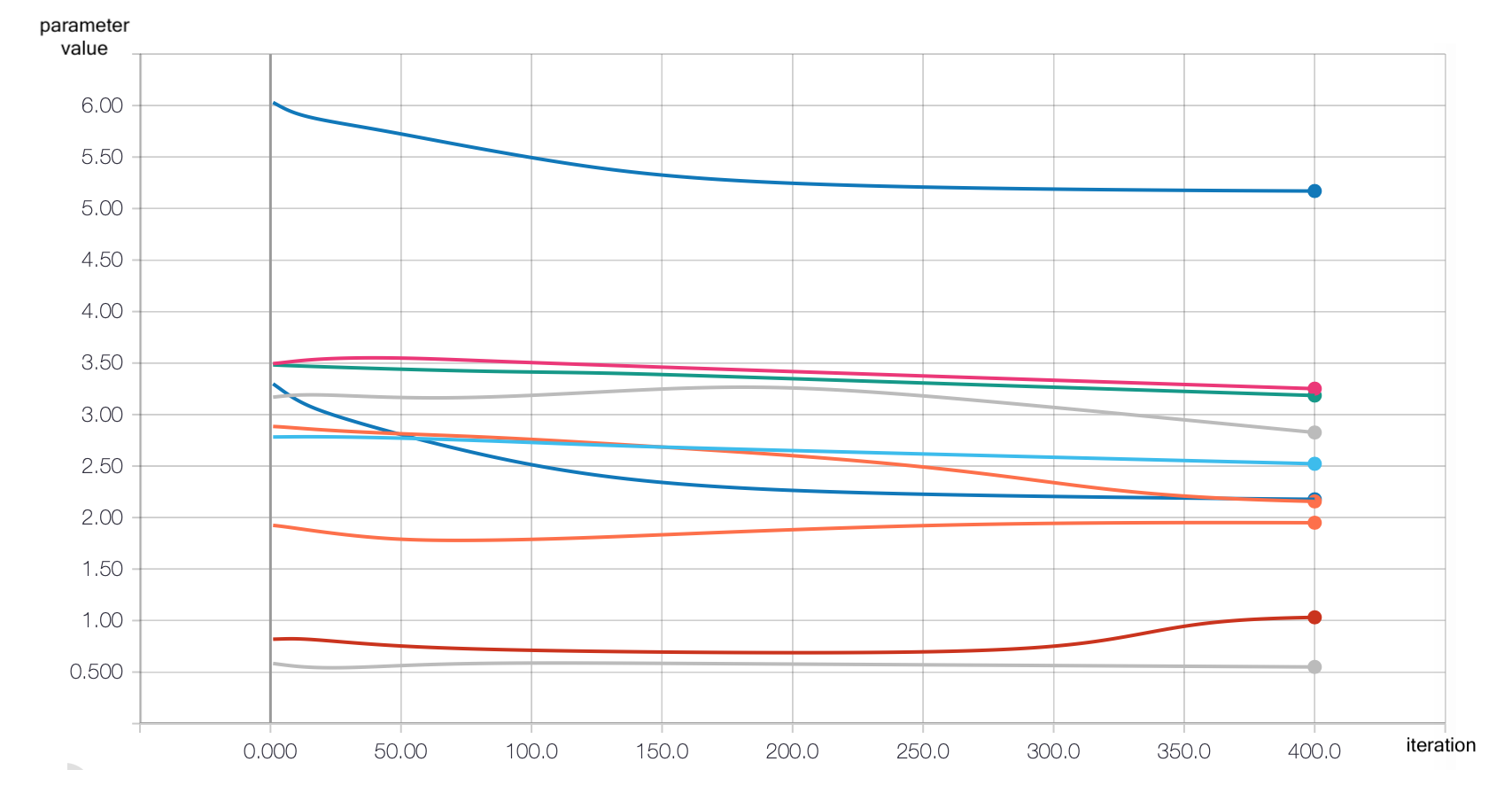} \\[\abovecaptionskip]
        \small B
    \end{tabular}
    
    \caption{Displacement gate: magnitude (A) converges towards specific values but we note that the phase (B) is more or less constant during the training.}
\label{fig:d_gate}
\end{figure}

\begin{figure}
    \centering
    \begin{tabular}{@{}c@{}}
        \includegraphics[width=1\linewidth,height=150pt]{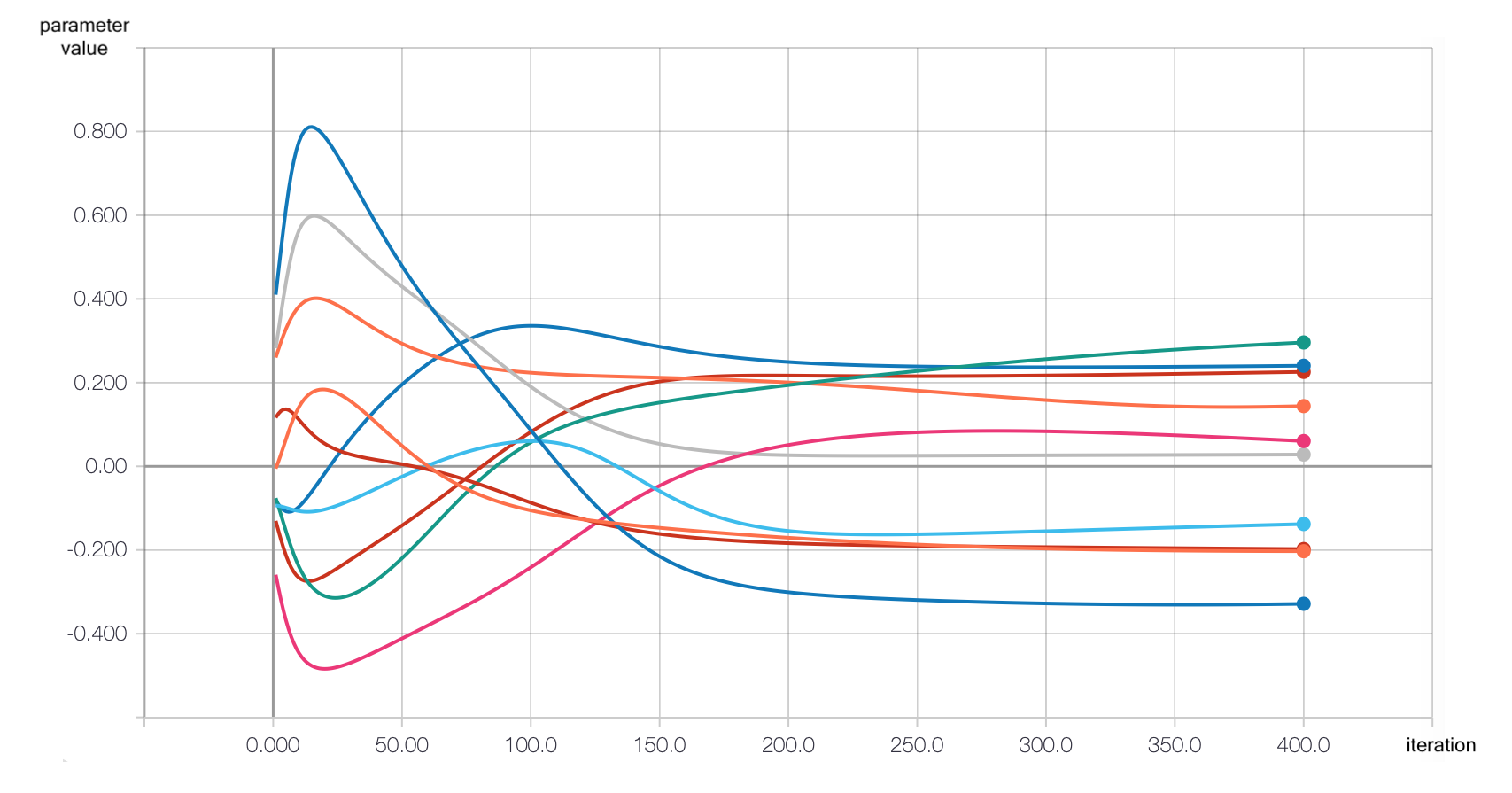} \\[\abovecaptionskip]
        \small A
    \end{tabular}
    
    \vspace{\floatsep}
    
    \begin{tabular}{@{}c@{}}
        \includegraphics[width=1\linewidth,height=150pt]{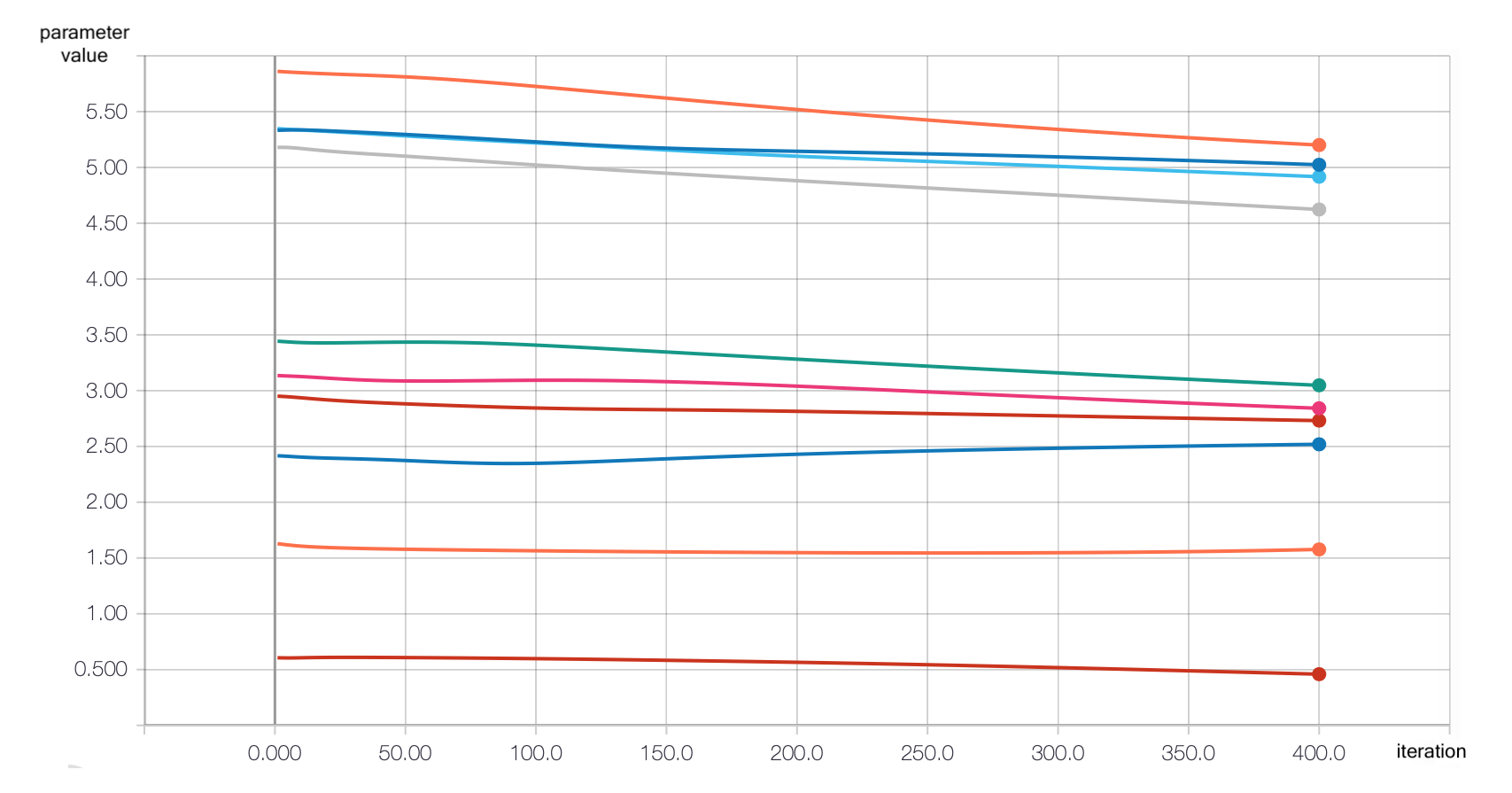} \\[\abovecaptionskip]
        \small B
    \end{tabular}
    
    \caption{Squeeze gate: the magnitude parameter (A) fails to converge towards specific values but we note that the phase (B) is more or less constant during the training.}
\label{fig:s_gate}
\end{figure}

\begin{figure}
    \centering
    \includegraphics[width=1\linewidth, height=150pt]{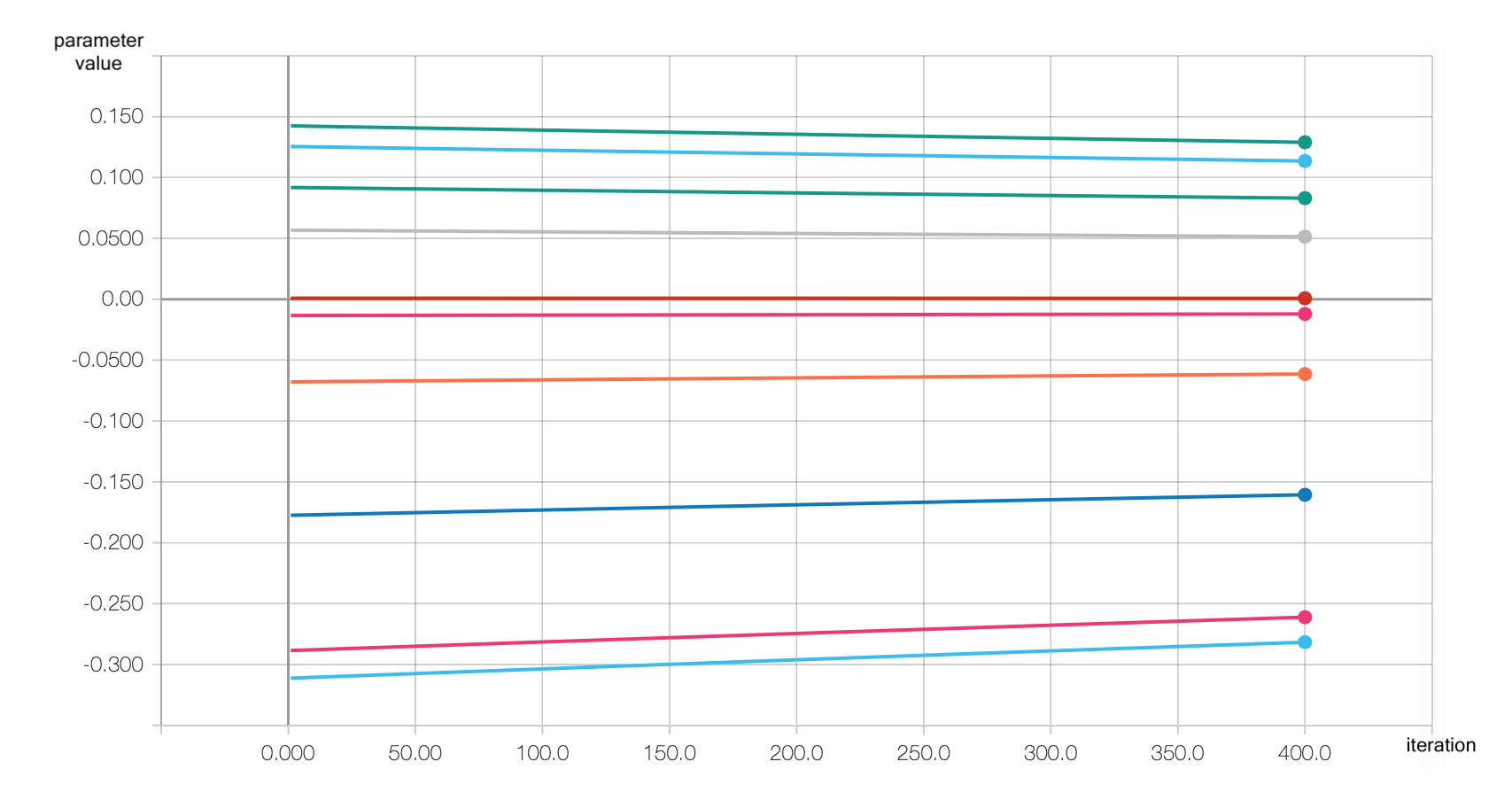}
    \caption{The Kerr gate parameter. The small change that can be observed on the plot comes from the regularization - in the absence of regularization, the parameter stays constant throughout the training.}
\label{fig:kerr_gate}
\end{figure}

\begin{figure}
    \centering
    \includegraphics[width=1\linewidth, height=140pt]{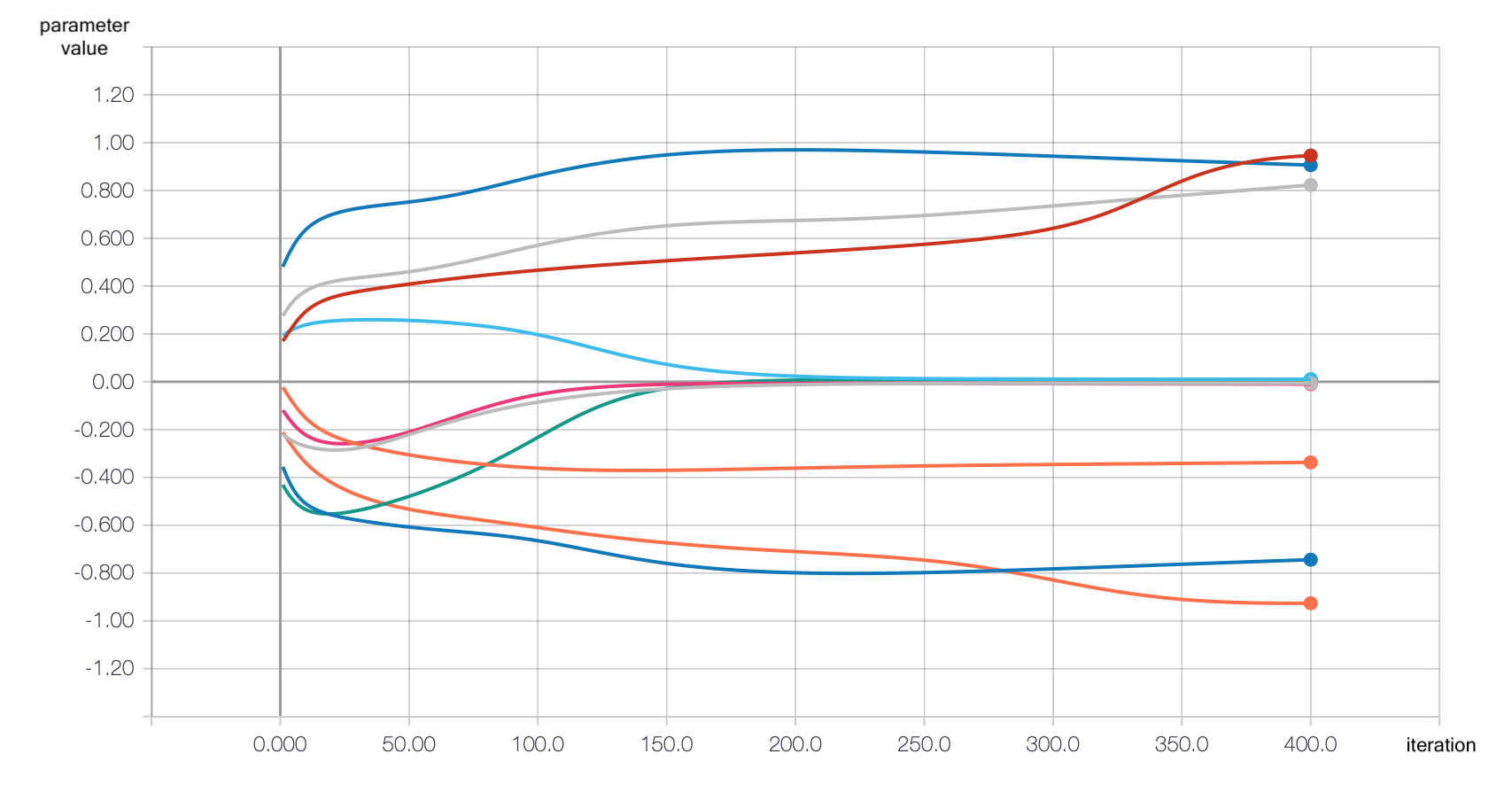}
    \caption{The cubic phase gate parameters converge to several different values during the optimization process.}
\label{fig:cubic_phase_gate}
\end{figure}

\begin{figure}
\centering
\begin{tabular}{@{}c@{}}
    \includegraphics[width=0.9\linewidth,height=130pt]{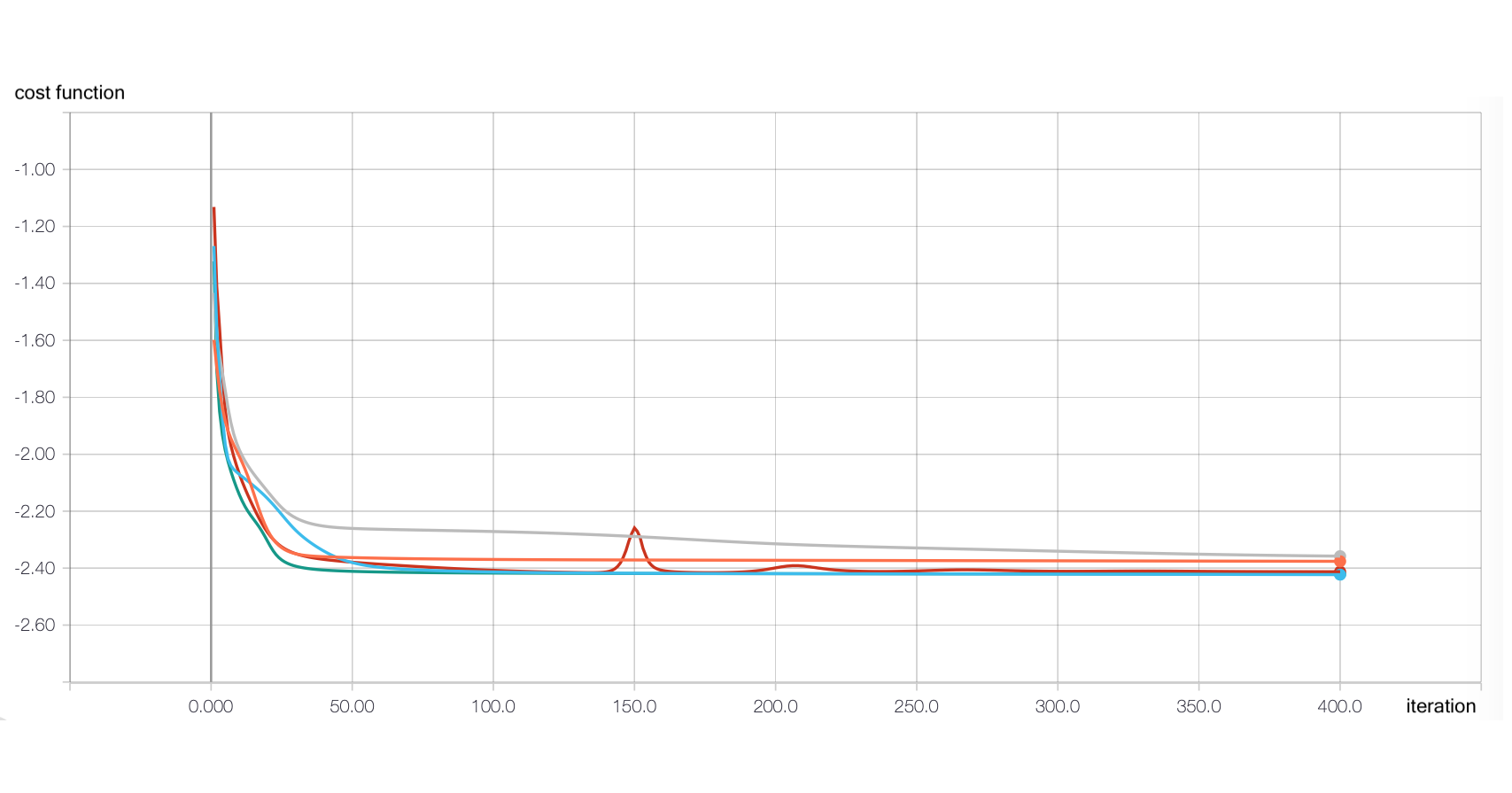} \\[\abovecaptionskip]
    \small A
\end{tabular}

\vspace{\floatsep}

\begin{tabular}{@{}c@{}}
    \includegraphics[width=0.9\linewidth,height=130pt]{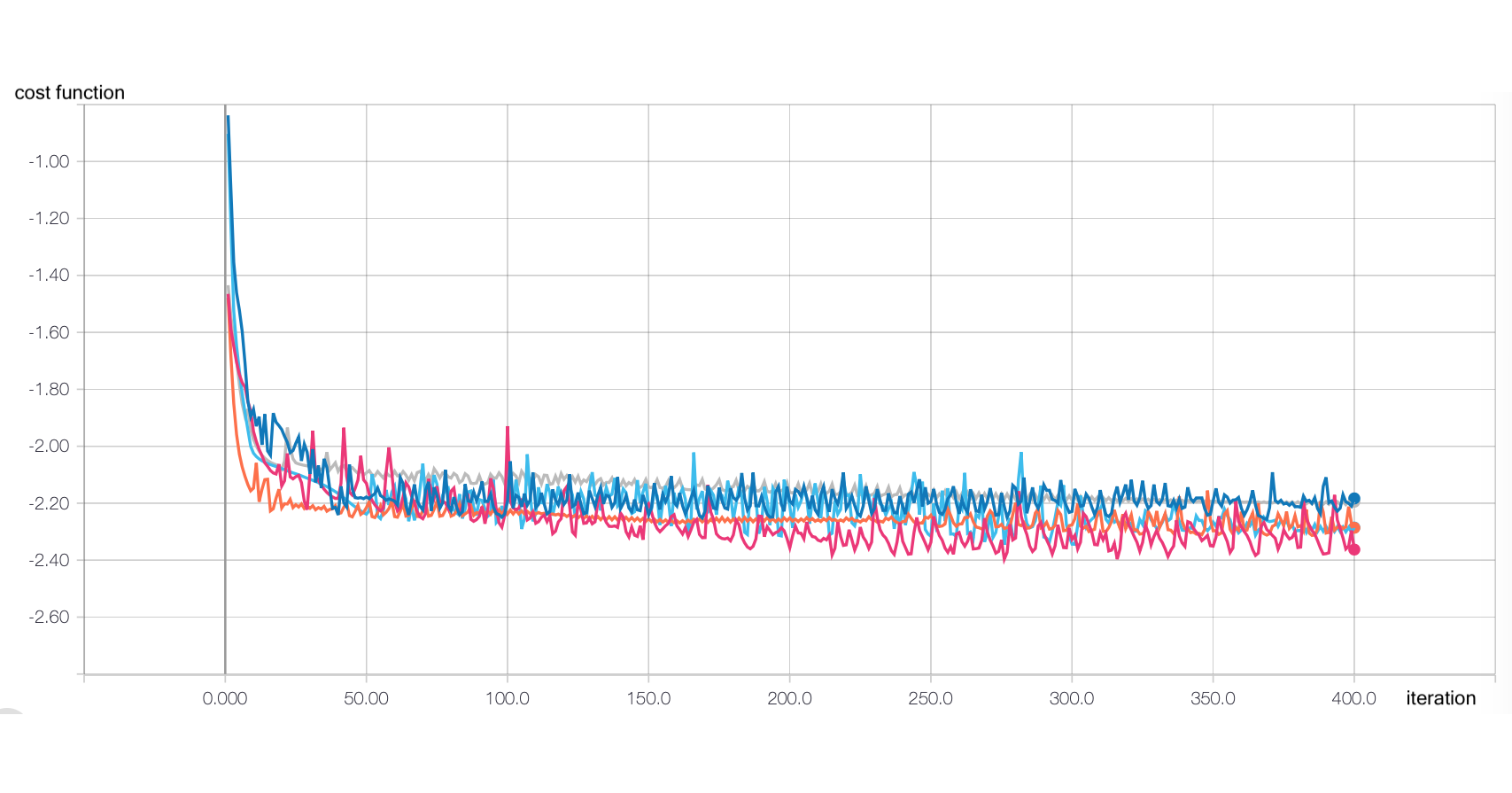} \\[\abovecaptionskip]
    \small B
\end{tabular}

\begin{tabular}{@{}c@{}}
    \includegraphics[width=0.9\linewidth,height=130pt]{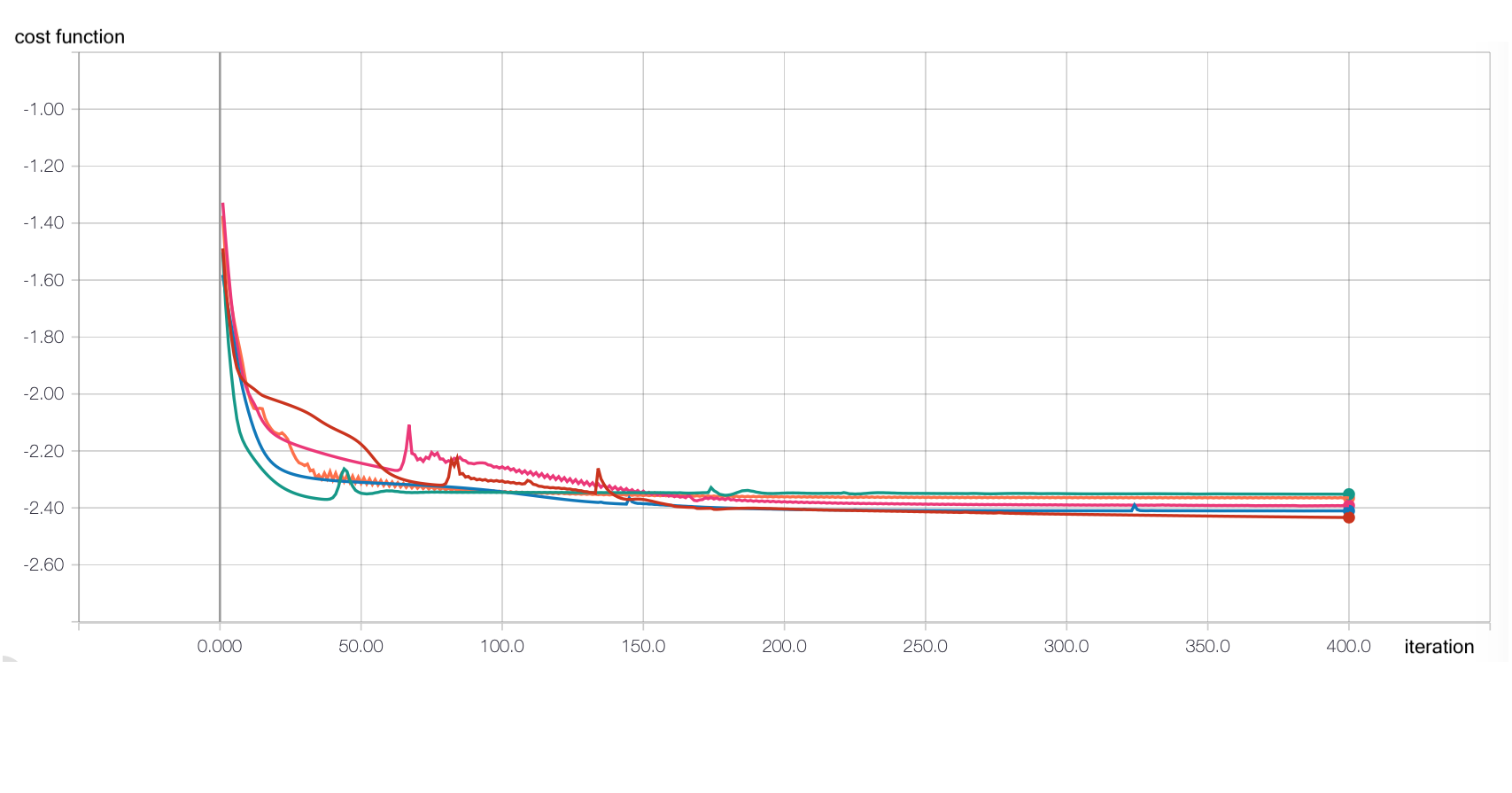} \\[\abovecaptionskip]
    \small C
\end{tabular}

\caption{Plots of the value of cost function as a function of number of iterations, for the machine learning approach with one layer. Each plot shows the results for 5 simulations. The type of non-Gaussian gates used for simulation are: none (A), Kerr gates (B), cubic phase gates (C).}
\label{fig:ml_results}
\end{figure}


\newpage



\begin{thebibliography}{9}
    
\bibitem{gaussian_quantum_information}
Weedbrook, Christian and Pirandola, Stefano and Garc\'{\i}a-Patr\'on, Ra\'ul and Cerf, Nicolas J. and Ralph, Timothy C. and Shapiro, Jeffrey H. and Lloyd, Seth "Gaussian Quantum Information".  {\em Rev. Mod. Phys.}. \textbf{84}, 621--669 (2012)
\url{https://link.aps.org/doi/10.1103/RevModPhys.84.621}

\bibitem{qaoa} Edward Farhi, Jeffrey Goldstone, "A Quantum Approximate Optimization Algorithm", \url{https://arxiv.org/abs/1509.04279}

\bibitem{crooks} Gavin E. Crooks, "Performance of the Quantum Approximate Optimization Algorithm on the Maximum Cut Problem", \url{http://arxiv.org/abs/1811.08419}

\bibitem{maxcut_d_wave}
Ryan Hamerly et al. "Experimental investigation of performance differences between coherent Ising machines and a quantum annealer". {\em Science Advances}. \textbf{5} (2019)
\url{https://advances.sciencemag.org/content/5/5/eaau0823}

\bibitem{var_alg}
Jarrod R. McClean, Jonathan Romero, Ryan Babbush, Al\'an Aspuru-Guzik, "The theory of variational hybrid quantum-classical algorithms"  {\em New Journal of Physics}. \textbf{18} (2016)
\url{https://iopscience.iop.org/article/10.1088/1367-2630/18/2/023023}

\bibitem{vqe}
Alberto Peruzzo, Jarrod McClean, Peter Shadbolt, Man-Hong Yung, Xiao-Qi Zhou Peter J. Love,4 Al\'an Aspuru-Guzik and Jeremy L. O'Brien, "A variational eigenvalue solver on a quantum processor" {\em Nature Communications}. \textbf{5} (2014)
\url{https://www.nature.com/articles/ncomms5213}

\bibitem{hadfield}
 Stuart Hadfield, Zhihui Wang, Bryan O'Gorman, Eleanor G. Rieffel, Davide Venturelli and Rupak Biswas, "From the Quantum Approximate Optimization Algorithm to a Quantum Alternating Operator Ansatz" {\em Algorithms}. \textbf{12} (2019)
\url{https://www.mdpi.com/1999-4893/12/2/34}

\bibitem{cv-qaoa} Guillaume Verdon, Juan Miguel Arrazola, Kamil Br\'adler and Nathan Killoran, "A Quantum Approximate Optimization Algorithm for continuous problems", \url{http://arxiv.org/abs/1902.00409}.

\bibitem{zhou} Leo Zhou, Sheng-Tao Wang, Soonwon Choi, Hannes Pichler and Mikhail D. Lukin "Quantum Approximate Optimization Algorithm: Performance, Mechanism, and Implementation on Near-Term Devices", \url{http://arxiv.org/abs/1812.01041}

\bibitem{gbs2}
Kamil Bradler, Pierre-Luc Dallaire-Demers, Patrick Rebentrost, Daiqin Su, and Christian Weedbrook, "Gaussian Boson Sampling for perfect matchings of arbitrary graphs" {\em Phys. Rev. A}. \textbf{98} (2018)
\url{https://link.aps.org/doi/10.1103/PhysRevA.98.032310}

\bibitem{cvqnn} Nathan Killoran, Thomas R. Bromley, Juan Miguel Arrazola, Maria Schuld,Nicol\'as Quesada, and Seth Lloyd, "Continuous-variable quantum neural networks", \url{https://arxiv.org/pdf/1806.06871}

\bibitem{gbs} Craig S. Hamilton, Regina Kruse, Linda Sansoni, Sonja Barkhofen, Christine Silberhorn, and Igor Jex, "Gaussian Boson Sampling", \url{https://arxiv.org/pdf/1612.01199}

\bibitem{sf} Nathan Killoran, Josh Izaac, Nicol\'as Quesada, Ville Bergholm, Matthew Amy, and Christian Weedbrook "Strawberry Fields:
A Software Platform for Photonic Quantum Computing", \url{https://arxiv.org/pdf/1804.03159}

\bibitem{qmlt} Maria Schuld and Josh Izaac. Xanadu Quantum Machine Learning Toolbox documentation. \url{https://qmlt.readthedocs.io}

\bibitem{takagi} Carlton M. Caves, "Polar decomposition, singular-value decomposition, and Autonne-Takagi factorization", \url{http://info.phys.unm.edu/~caves/courses/qinfo-s17/lectures/polarsingularAutonne.pdf}
    
\end{thebibliography}
\end{document}